\documentclass[fleqn,usenatbib]{mnras}
\usepackage{newtxtext,newtxmath}
\usepackage[T1]{fontenc}
\DeclareRobustCommand{\VAN}[3]{#2}
\let\VANthebibliography\thebibliography
\def\thebibliography{\DeclareRobustCommand{\VAN}[3]{##3}\VANthebibliography}
\usepackage{graphicx}	% Including figure files
\usepackage{amsmath}	% Advanced maths commands
\usepackage{xcolor}

\newcommand{\mpc}{{{\, \rm Mpc}}}

\title[Stellar halos in the cosmic web]{Spatial and velocity anisotropies of stellar halos across cosmic web environments: Insights from IllustrisTNG simulation}
\author[A. Mondal et al.]{
Amit Mondal$^1$\thanks{E-mail:amitmondal.bwn95@gmail.com},
Biswajit Pandey$^1$\thanks{E-mail:biswap@visva-bharati.ac.in} and
Anindita Nandi $^1$\thanks{E-mail:anindita.nandi96@gmail.com} \\
$^1$Department of Physics, Visva-Bharati University, Santiniketan, 731235, West Bengal, India}

\date{Accepted XXX. Received YYY; in original form ZZZ}
\pubyear{\the\year{}}
\begin{document}
\label{firstpage}
\pagerange{\pageref{firstpage}--\pageref{lastpage}}
\maketitle

% Abstract of the paper

\begin{abstract}
  The role of large-scale environment in shaping the structural and kinematic properties of stellar halos remains an open question. We investigate whether the cosmic web environments affect the spatial and velocity anisotropies of stellar halos in Milky Way-mass galaxies. Using high-resolution data from the TNG50 simulation, we analyze 29 stellar halos from each environments: sheets, fillaments and clusters and quantify their spatial and kinematic anisotropies as a function of halo-centric radius. We find that stellar halos across all environments generally exhibit increasing spatial anisotropy with radius, with fluctuations corresponding to bound substructures. The velocity anisotropy profiles show radially dominated orbits on average, but also display significant local variation, including tangentially dominated regions. However, no statistically significant differences are observed in the mean spatial or velocity anisotropy profiles across environments, for either the total stellar halo population or for the in situ and ex situ components individually. The large scatter within each environment suggests that the formation of stellar halos is primarily driven by stochastic, small-scale processes such as satellite merger histories, rather than the large-scale geometry of the cosmic web. Our results imply that, at fixed halo mass, the influence of cosmic web environment on the structure of stellar halo is weak or highly non-deterministic. Possible environmental effects may be more prominent at higher masses where accretion is more anisotropic. Exploring this regime will require simulations with both larger volume and higher resolution.
\end{abstract}


\section{Introduction}

Stellar halos are among the most overlooked components of galaxies, despite their importance in tracing galactic history. They contain only $\sim 2\%$ of a galaxy's stellar mass \citep{deason19}, yet they offer a unique window into the formation and evolution of galaxies over cosmic time. Often dominated by metal-poor and old stars \citep{beers05, frebel15}, stellar halos can also exhibit a wide range of metallicities and ages, reflecting the diversity of galaxy assembly histories \citep{ibata01, brown08, gilbert14, dsouza18, harmsen17, harmsen23}. As the outermost and most diffuse component, the stellar halos preserve long-lived signatures of past accretion events and merger histories, making them a crucial probe of hierarchical galaxy formation. Early work by \citet{eggen62} attempted to reconstruct the formation history of the Milky Way’s stellar halo, proposing that it formed through the rapid collapse of a large, ancient gas cloud. However, subsequent studies \citep{searle78, helmi99} have shown that the stellar halo more likely formed through a series of accretion and merger events, consistent with the hierarchical structure formation predicted by the $\Lambda$CDM model. Stellar halos are built up via dissipationless processes, characterized by the chaotic and gradual merging of smaller subsystems. This assembly history results in significant spatial variation in the density and structure of the stellar population across the halo. The inner regions of the stellar halo are denser and characterized by shorter dynamical timescales compared to the more diffuse outer halo. Consequently, the debris from past accretion events in the inner halo becomes spatially well mixed, leading to a smoother and more homogeneous stellar distribution. In contrast, the outer halo, with its longer dynamical timescales, retains prominent substructures including the extended tidal streams. This distinction explains the abundance of coherent stellar streams observed in the outer halo of the Milky Way. Notable examples include the Sagittarius tidal stream \citep{ibata94, ivezic00, yanny00}, the low-latitude stream \citep{yanny03, ibata03}, and the Orphan Stream \citep{grillmair06, belokurov07, newberg10}, among many others that trace the remnants of past accretion events. Observations and simulations indicate that the stellar density profile approximately follows a power-law decline, scaling as $r^{-3}$ in the inner regions and becoming progressively steeper in the outer halo \citep{bell08, deason11, slater16}.

Understanding the different components of stellar halo and their properties would provide crucial insight into the hierarchical assembly of the galaxy over cosmic time. The stellar halo population can be broadly divided into two components based on their origin: in situ and ex situ. The in situ component consists of stars that formed within the host galaxy itself, reflecting its internal evolutionary processes. In contrast, the ex situ component comprises stars accreted from disrupted satellite galaxies or star clusters, offering insight into the galaxy’s merger and accretion history. A detailed study of in situ stellar halo formation in Milky Way-mass galaxies is presented in \citet{cooper15}, who analyzed stellar halos formed in smoothed particle hydrodynamics (SPH) simulations. The corresponding dark matter only versions of these simulations were part of the Aquarius Project \citep{springel08}. In their simulations, in situ stars dominate the stellar halo out to approximately 20 kpc, contributing about $30\%-40\%$ of the total halo mass. Their key finding is that the hierarchical assembly of structure drives the formation of both ex situ and in situ halo components, highlighting the interconnected nature of internal and external processes in shaping stellar halos.

The standard model of structure formation posits that galaxies form at the centers of dark matter halos \citep{white78, blumenthal84, white91, navarro97, springel05}. In this paradigm, the galaxy along with its stellar halo evolve within the gravitational potential of their surrounding dark matter halo. The galaxies are thus intricately connected to their host halos \citep{wechsler18}. The galaxies along with their stellar halos, and dark matter halos are embedded within the large-scale cosmic web, which provides the evolving environment for structure formation. Numerous observational studies indicate that tidal forces exerted by the large-scale structure (LSS) can affect the spin and orientation of galaxies \citep{lambas88, lee07, hirv17}. Specifically, galaxies located in filaments tend to have their major axes aligned along the filament direction, while those in sheets often align within the plane of the sheet \citep{jones10, zhang13, tempel13a, tempel13b, zhang15, chen19, krolewski19}. Similarly, cosmological simulations have shown that dark matter halos exhibit preferred alignments relative to the geometry of their host cosmic web environment \citep{zhang09, libeskind12, veena18}.

A study using the Illustris simulation found no correlation between stellar halo fraction and environment \citep{elias18}. However, the stellar halo fraction alone does not capture the detailed internal structure of stellar halos. In our work, we aim to explore how the environment influences the spatial and kinematic properties of stellar halos. It is also worth noting that the earlier study used local density as a proxy for environment. While useful, local density does not fully represent the environment of a galaxy. Galaxies are part of the cosmic web, which is an interconnected complex network of galaxies comprising of sheets, filaments and nodes surrounded by vast empty regions \citep{gregory78,joeveer78,einasto80,zeldovich82,einasto84}. Studies with N-body simulations reveal the dynamic nature of matter flow within the cosmic web \citep{arag10, cautun14, ramachandra, wang24}, showing a coherent progression of mass from voids into walls, then from walls into filaments, and ultimately toward the dense cluster. The geometry of the cosmic web is known to exert a distinct influence on the assembly history of galaxies by guiding the flow of matter and angular momentum onto halos \citep{dubois14, laigle15}. The cosmic web is also known to play a fundamental role in shaping many galaxy properties. Early evidence for environmental effects was provided by Dressler \citep{dressler80}, and subsequent studies have demonstrated that galaxy luminosity, colour, morphology, and star formation rates are strongly affected by the filamentarity and connectivity of the large-scale structure \citep{pandey05, pandey06, pandey08, galarraga23}. Both observations and simulations consistently show that galaxies residing closer to filaments tend to be more massive, redder, and more quenched, even when controlling for halo mass and local density \citep{malavasi17, kuutma17, kraljic18, pandey20, singh20}. Furthermore, several studies have confirmed that stellar mass, colour, star formation activity, and spin alignments are modulated by the anisotropic nature of the cosmic web \citep{veena18, das23, hoosain24, nandi24}. Filaments are known to be the largest known coherent patterns in the cosmic web \citep{bharad04, pandey10, pandey11, sarkar23}. They host a significant fraction ($40\%-50\%$) of the baryonic matter in the form of the Warm-Hot Intergalactic Medium (WHIM) \citep{tuominen21, galarraga21}. The diffuse WHIM serves as a vast reservoir that can accrete onto galaxies over time, fueling their growth and influencing their star formation activity. High-resolution zoom-in hydrodynamical simulations by \citet{liao19} demonstrate that gas accretion onto halos embedded in filaments is strongly anisotropic. This directional inflow leads to enhanced baryon and stellar mass fractions in filament-hosted halos compared to those in more isolated, field environments. Complementary studies using both dark matter only and hydrodynamical simulations \citep{veena18, veena19} further reveal a mass-dependent accretion pattern, where low-mass halos tend to accrete material perpendicular to the filament axis, while high-mass halos accrete predominantly along the axis of the filament. A study \citep{kraljic20} with the SIMBA simulation \citep{dave19} support this trend, showing that galaxies within filaments exhibit HI content signatures consistent with anisotropic gas inflow. In this context, stellar halo acts as a fossil record of the formation and evolutionary history of its host galaxy, preserving signatures of ancient mergers and accretion events. These diffuse and extended components are sensitive to the orbital and spatial characteristics of infalling satellites \citep{bullock05, cooper10, pillepich15, monachesi19}. Since accretion is often anisotropic and modulated by large-scale filaments and walls, one may expect that the structure and kinematics of stellar halos might reflect the cosmic web geometry in which a galaxy resides. However, it remains an open question whether the structural and kinematic anisotropies of stellar halos \citep{mondal24}, which directly encode the mode and direction of mass accretion, vary systematically with cosmic web environment.

Studying the spatial and velocity anisotropies of these components across different large-scale environments such as sheets, filaments, and clusters can provide insight into whether the cosmic web influences not just the mass assembly of halos, but also their detailed internal structure. Previous studies have suggested that accretion is anisotropic and environment dependent, particularly at high masses \citep{dubois14, pillepich18a}, raising the possibility that stellar halos in different environments might show systematic difference in their kinematic or spatial signatures. We investigate this possibility by analyzing the spatial and velocity anisotropy of in situ, ex situ, and combined stellar halo populations across different cosmic web environments.

To address this, we analyze a suite of Milky Way-mass galaxies drawn from the IllustrisTNG simulations \citep{nelson19}, classified according to their position in the cosmic web. By focusing on galaxies in different geometric environments such as sheets, filaments, and clusters, we aim to isolate the influence of large-scale structure on the spatial and velocity anisotropies of their stellar halos. We analyze stellar halos from the TNG50 Milky Way/Andromeda-like galaxy catalogue \citep{pillepich24}. A Hessian based method \citep{hahn07} is employed to classify these stellar halos according to their host environments in the cosmic web. We analyze the spatial anisotropies \citep{pandey22} and the velocity anisotropies \citep{binney80} of the stellar halos across the different cosmic web environments. We also separately examine the spatial and kinematic properties of the in situ and ex situ components of stellar halos across various cosmic web environments and compare their characteristics.

Our paper is organized as follows. In Section 2, we describe the simulation data and our classification of the cosmic web. We also discuss the methodology for quantifying anisotropies in stellar halos in Section 2. We report our main results in Section 3, and present our conclusions in Section 4.

%%%%%%%%%%%%%%%%%%%%%%%%%%%%%%%%%%%%%%%%%%%%%%%%%%%%%%%%%%%
 
\section{Data and method of analysis} 
\subsection{Data}

The IllustrisTNG simulation suite\footnote{\url{https://www.tng-project.org/}} \citep{nelson18, nelson19, springel18, pillepich18a, marinacci18, naiman18} provides a state of the art framework for modeling galaxy formation and evolution within a cosmological context. Utilizing advanced numerical methods, it captures the complex interplay between dark matter, gas, and stars across a large cosmological volume. The simulation incorporates detailed physical processes including gas dynamics, star formation, supernova feedback, and black hole growth to offer comprehensive insights into the mechanisms driving galaxy evolution. IllustrisTNG simulations are built on the moving-mesh code AREPO \citep{springel10, weinberger20}. It extends and improves upon the original Illustris project \citep{vogelsberger13, vogelsberger14, genel14} by incorporating updated physics models and enhanced numerical techniques. The TNG suite comprises three simulation volumes with side lengths of approximately 50, 100, and 300 Mpc which are designated as TNG50, TNG100, and TNG300, respectively. Each volume is available at multiple resolution levels (e.g., TNG100-1, TNG100-2, TNG100-3), with TNG50 providing four distinct resolution outputs. All these simulations start at a redshift of $z = 127$ and adopt cosmological parameters consistent with the Planck 2015 results \citep{planck16}.

For the present analysis, we use $z=0$ snapshot of the high-resolution TNG50-1 simulation, which provides the matter distributions in a cubic box of side length 51.7 comoving\mpc. TNG50 resolves baryonic matter with a mass resolution of $8.5 \times 10^4\, M_\odot$ and dark matter with a mass resolution of $4.5 \times 10^5\, M_\odot$, while achieving a spatial resolution of approximately $100-140$ parsecs for star-forming ISM gas \citep{pillepich19}.

In our analysis, we consider the TNG50 Milky Way/Andromeda-like galaxy catalogue \citep{pillepich24}, which includes 198 MW/M31-like galaxies at $z=0$. The selection criteria for this sample are the following. 

(1) The galaxy stellar mass should be in the range $M_*(<30 \, \text{kpc}) = 10^{10.5-11.2} M_\odot$. 

(2) The galaxy must have disk like morphology and spiral arms.

(3) There should be no other galaxy with stellar mass $\geq 10^{10.5} M_\odot$ within a distance of 500 kpc from the galaxy and the total mass of the host halo is smaller than $M_{200} (\text{host}) < 10^{13} M_\odot$.

This datasets provide all gas, star, and dark-matter particles within $\pm 400$ comoving kpc of the galaxy center, without removing satellites or substructures. As a result, our stellar-halo sample includes stars bound to the central galaxy as well as those still bound to surviving satellites.
The database of TNG50-1 simulation provides star particles pre-classified as in situ or ex situ, based on their origin within or outside the main progenitor halo. Dividing halo stars into in situ and ex situ components provides a physically meaningful decomposition that reflects different phases of halo assembly \citep{zolotov09, pillepich15}. The ex situ stars, in particular, are expected to trace the hierarchical growth of galaxies through satellite accretion and tidal disruption, often along preferred directions set by the cosmic web. In contrast, in situ stars formed through internal star formation or dynamical heating may exhibit more isotropic structures influenced by baryonic feedback, disk potential, or past merger-induced perturbations \citep{villa08, zolotov09, jean17, bonaca17}.

\subsection{Method of analysis}

\begin{figure*}
\resizebox{15cm}{5cm}{\rotatebox{0}{\includegraphics{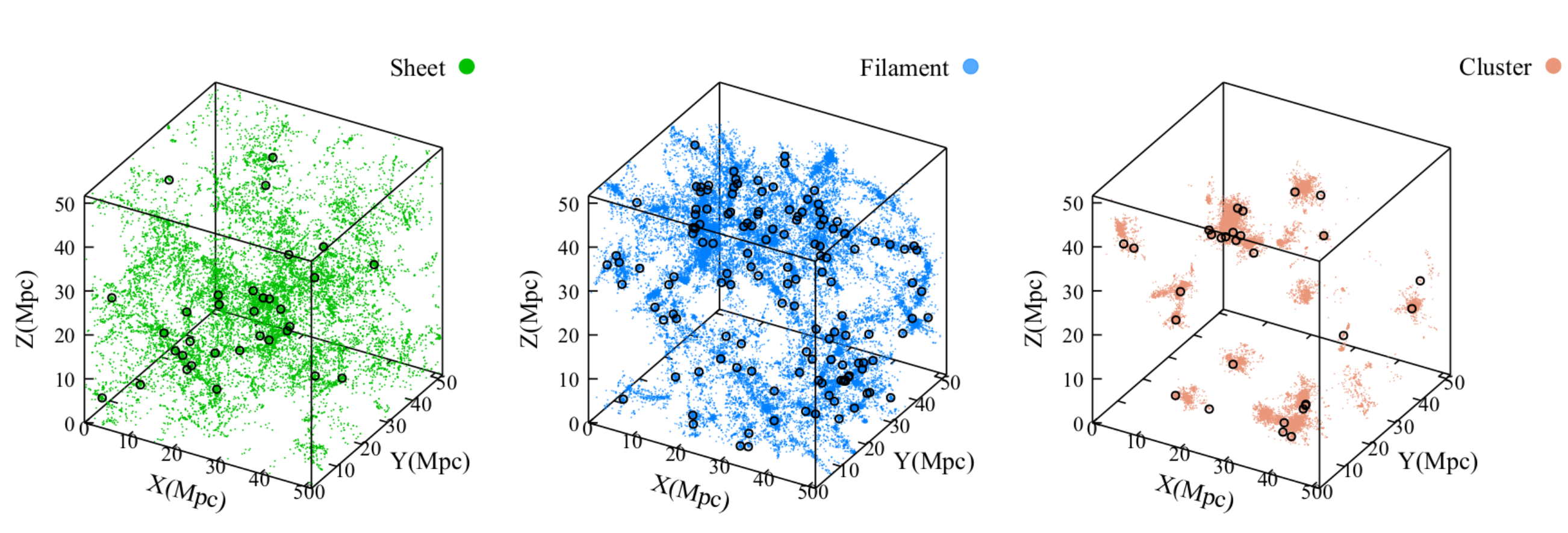}}}
\caption{The three panels show the spatial distribution of TNG50 galaxies, color-coded by cosmic web environment, along with the positions of 198 stellar halos from Milky Way/Andromeda-like galaxy catalogue. The circles representing stellar halos are visually enlarged for clarity and their sizes do not reflect the actual physical extent of the halos.}
\label{fig:env_combine}
\end{figure*}

\subsubsection{Quantifying the cosmic web environments of stellar halos}
\label{sec:cosmic_web_classification}

We identify the morphological components of the cosmic web (voids, sheets, filaments, and clusters) by employing a Hessian-based method that utilizes the eigenvalues of the deformation tensor derived from the gravitational potential \citep{hahn07, forero09}.

The deformation tensor $T_{ij}$ is defined as the Hessian of the gravitational potential $\Phi(\mathbf{x})$.
\begin{equation}
T_{\alpha\beta} = \frac{\partial^2 \Phi}{\partial x^\alpha \partial x^\beta}, \quad \alpha,\beta \in \{1,2,3\}.
\label{eq:hessian}
\end{equation} 

The gravitational potential $\Phi$ is related to the density contrast field $\delta(\mathbf{x})$ via the Poisson equation,

\begin{equation}
\nabla^{2}\phi \equiv \delta
\label{eq:poisson}
\end{equation} 
where, $\delta=\frac{\rho - \bar{\rho}}{ \bar{\rho}}$ is the fractional density contrast.

To begin, we transform the discrete galaxy distribution into a continuous density field using the Cloud in Cell (CIC) interpolation method on a three-dimensional grid of size $128^3$. This discretized density field is then Fourier transformed and smoothed by applying an isotropic Gaussian window function. Specifically, we adopt a smoothing length of $3.23\,\mathrm{Mpc}$, chosen to ensure that only large-scale structures are retained. This scale is selected based on our focus on the characterization of large-scale cosmic environments, while suppressing small-scale fluctuations that may not be relevant to our analysis.

We calculate the gravitational potential corresponding to the smoothed density field in the Fourier space.

\begin{equation}
  \hat{\Phi}(\mathbf{k}) = \hat{G}(\mathbf{k}) \, \hat{\rho}(\mathbf{k})
  \label{eq:phik}
\end{equation}

Here, $\hat{\Phi}(\mathbf{k})$ is the Fourier transform of the gravitational potential $\Phi(\mathbf{x})$, and $\hat{\rho}(\mathbf{k})$ is the Fourier-transformed smoothed density contrast field. The term $\hat{G}(\mathbf{k})$ represents the Green's function of the Laplacian operator in Fourier space. Once $\hat{\Phi}(\mathbf{k})$ is computed, we apply an inverse Fourier transform to obtain the gravitational potential $\Phi(\mathbf{x})$ in real space. Finally, we evaluate the deformation tensor using numerical differentiation on the potential field.

We compute the three eigenvalues $\lambda_1 > \lambda_2 > \lambda_3$ of $T_{ij}$ at each grid point. Based on the signs of these eigenvalues, we classify the local environment into one of the four cosmic web environments,

\begin{itemize}
    \item Void: $\lambda_1 < 0$, $\lambda_2 < 0$, $\lambda_3 < 0$
    \item Sheet: $\lambda_1 > 0$, $\lambda_2 < 0$, $\lambda_3 < 0$
    \item Filament: $\lambda_1 > 0$, $\lambda_2 > 0$, $\lambda_3 < 0$
    \item Cluster: $\lambda_1 > 0$, $\lambda_2 > 0$, $\lambda_3 > 0$
\end{itemize}

Each galaxy is assigned the classification corresponding to its grid location.

\begin{table}
	\centering
	\caption{This table presents the number of galaxies from the MW/M31-like catalogue (total of 198 galaxies) across different morphological environments.}
	\label{tab:count_environment}
	\begin{tabular}{l c} % two columns: left and center aligned
		\hline
		Morphological Environment & Number of Galaxies \\
		\hline
		Void     & 0   \\
		Sheet    & 33  \\
		Filament & 136 \\
		Cluster  & 29  \\
		\hline
	\end{tabular}
\end{table}

We tabulate the number of galaxies from the TNG50 Milky Way/Andromeda-like galaxy catalogue that are identified in various morphological environments in \autoref{tab:count_environment}. We also show the spatial distribution of TNG50 galaxies, classified by cosmic web environment, alongside the locations of the 198 stellar halos from the MW/M31-like catalogue in the separate panels of \autoref{fig:env_combine}.

\subsubsection{Separating the stellar halo components of the galaxies in MW/M31-like catalogue}

We ensure an equal number of stellar halos from each environment to enable a fair comparison of spatial and velocity anisotropies across different cosmic web environments. In our sample, we identify 33 stellar halos residing in sheets, 136 in filaments, and 29 in clusters \autoref{tab:count_environment}. Given that clusters contain the fewest halos, we adopt 29 as the baseline and randomly select 29 stellar halos from both the sheet and filament environments. This approach eliminates potential biases arising from unequal sample sizes and allows for a balanced and statistically meaningful comparison of the structural and kinematic properties of stellar halos across sheets, filaments, and clusters. 

The different components of a galaxy can be separated according to their morphology such as discs and spheroidal components (e.g., bulges and stellar halos). The galaxies are composed of gas cells, dark matter particles, stars, wind particles, and supermassive black holes (SMBHs). For the present work, we focus exclusively on the stellar component of the stellar halo. The data structure of the TNG50 Milky Way/Andromeda-like galaxy catalogue is designed to allow easy identification of stellar particles associated with individual galaxies. Our goal is to isolate the stellar halo population from the total stellar content of each galaxy for detailed analysis.

In our analysis, we use the \textit{RotatedCoordinates} $(x, y, z)$ and \textit{RotatedVelocities} $(v_x, v_y, v_z)$ of halo stars. These quantities represent the spatial positions and velocities of stars in a reference frame aligned with the main galaxy. The \textit{RotatedCoordinates} describe the stellar positions relative to the intrinsic coordinate system of the galaxy and are obtained in two steps: (1) a transformation shifts the coordinate origin to the center of the main galaxy, and (2) a rotation is applied by diagonalizing the moment of inertia tensor of either the stellar or star-forming gas particles within one or two times the stellar half-mass radius \citep{pillepich24}. This rotation aligns the minor axis of the galaxy with the $z$-axis. Similarly, the \textit{RotatedVelocities} represent the stellar velocities in a frame where the main galaxy is at rest. These are also computed in two steps: (1) the peculiar velocity of the galaxy through the simulation volume is subtracted from each stellar velocity, and (2) the same rotational transformation used for the coordinates is applied. A detailed descriptions can be found in \url{https://www.tng-project.org/data/milkyway+andromeda}.

To separate the halo stars from the disc star and bulge star we use the method proposed by \citet{scannapieco09}. They defined the circularity parameter of a star as, $\epsilon = \frac{j_z}{j_{\text{circ}}}$, where $j_z$ is the angular momentum of the star along z-direction, which is actually perpendicular to the direction of the galactic disc and $j_{\text{circ}}$ is the angular momentum of the star if the orbit of the star is circular with the same radius. $j_{\text{circ}}$ is defined as $j_{\text{circ}} = r\,v_{\text{circ}}$, where $v_{\text{circ}}=\sqrt{\frac{GM(\leq r)}{r}}$ and $M(\leq r)$ is the mass enclosed within $r$ by taking account of all the matter components (i.e. gas cells, DM particles, stars, wind particles and SMBHs). We calculate the circularity parameter of each of the stars in a galaxy and only extract the stars with  $\epsilon < 0.5$ and  $r>3.5$ kpc \citep{zhu22}. This approach enables the effective removal of most stars associated with the disc and bulge components of the galaxy.

\begin{table}
    \centering
    \scriptsize
    \caption{This table shows the counts of stars in 29 stellar halos chosen for this analysis from different morphological environments.}
    \label{tab:counts_stellar_halos}
    \begin{tabular}{rr rr rr}
        \hline
        \multicolumn{2}{c}{\textbf{Sheet}} & 
        \multicolumn{2}{c}{\textbf{Filament}} & 
        \multicolumn{2}{c}{\textbf{Cluster}} \\
        \textbf{Subhalo ID} & \textbf{Count} & 
        \textbf{Subhalo ID} & \textbf{Count} & 
        \textbf{Subhalo ID} & \textbf{Count} \\
        \hline
        422754 & 1053099 & 342447 & 1441680 & 358609 & 797874 \\
        428177 & 1059334 & 372754 & 1576167 & 424288 & 1308611 \\
        430864 & 1803847 & 372755 & 849235  & 429471 & 491683 \\
        441709 & 1300265 & 388544 & 2574495 & 432106 & 901069 \\
        452031 & 778304  & 392277 & 116458  & 436932 & 696741 \\
        464163 & 846578  & 400973 & 1620446 & 461785 & 1046732 \\
        473329 & 1369513 & 400974 & 418242  & 471996 & 912845 \\
        474008 & 617313  & 402555 & 1050301 & 475619 & 1093882 \\
        482889 & 1203994 & 411449 & 2183472 & 479290 & 1311996 \\
        492876 & 1102711 & 414917 & 1811993 & 485056 & 922495 \\
        494709 & 303044  & 414918 & 385706  & 489206 & 410540 \\
        504559 & 396378  & 416713 & 1064904 & 490079 & 652014 \\
        510585 & 315229  & 419618 & 544866  & 505586 & 324928 \\
        512425 & 506196  & 421555 & 1450282 & 507784 & 321521 \\
        516101 & 328908  & 425719 & 1682024 & 511920 & 409626 \\
        520885 & 553850  & 427211 & 1111440 & 513845 & 926283 \\
        523548 & 832914  & 433289 & 1106806 & 522530 & 318392 \\
        528322 & 226214  & 435752 & 1557163 & 522983 & 325968 \\
        531320 & 317136  & 438148 & 2526214 & 527309 & 458007 \\
        532301 & 355123  & 439099 & 1660880 & 530330 & 617394 \\
        534628 & 295316  & 440407 & 1052761 & 531910 & 165487 \\
        538370 & 418728  & 443049 & 963804  & 535774 & 358098 \\
        542662 & 149927  & 445626 & 1666854 & 539333 & 91871  \\
        543729 & 659887  & 446665 & 420185  & 543114 & 416476 \\
        547844 & 504772  & 447914 & 553093  & 554798 & 516038 \\
        549516 & 120190  & 448830 & 1691399 & 559386 & 177927 \\
        550149 & 150262  & 452978 & 1625246 & 565089 & 426755 \\
        552581 & 181018  & 454171 & 120329  & 569251 & 470031 \\
        554523 & 75497   & 454172 & 444532  & 613192 & 214062 \\
        \hline
    \end{tabular}
\end{table}

\subsubsection{Whole-sky anisotropy parameter}

We calculate the Whole-sky anisotropy parameter as a function of the radial distance from the centre 29 stellar halos from each cosmic web environment. The whole-sky anisotropy parameter \citep{pandey16, pandey22}, which quantifies spatial anisotropy, is based on information entropy \citep{shannon48}. The idea of the information entropy was originally introduced by Claude Shannon \citep{shannon48}, which quantifies the average amount of information required to describe a random variable.

The Shannon entropy $H(x)$ of a discrete random variable $x$ with $n$ possible outcomes $\{x_i\}_{i=1}^n$ is defined as,

\begin{equation}
H(x)= - \sum_{i=1}^{n}  p(x_{i})  \log  p(x_{i}),
\label{eq:shannon}
\end{equation}

where $p(x_i)$ is the probability of the $i^{th}$ outcome, and $-\log p(x_i)$ represents the information content of that outcome. Since $0 < p(x_i) \leq 1$, the entropy $H(x)$ measures the expected information required to specify the state of $x$. A perfectly certain outcome (i.e., $p=1$) carries no information, while maximum uncertainty (i.e., low $p$) yields higher entropy.

In our case, the random variable $x$ corresponds to the angular position of a randomly selected star particle within a distance $r$ from the centre of the stellar halo. For each radial distance $r$, the full $4\pi$ steradian solid angle is pixelated into equal-area elements using the Hierarchical Equal Area isoLatitude Pixelization (HEALPix) scheme \citep{gorski05}. The number of angular pixels is determined by the HEALPix resolution parameter $N_{\text{side}}$, such that,

\begin{equation}
N_{\text{pix}} = 12 \, N_{\text{side}}^2.
\label{eq:npix}
\end{equation}

Each pixel thus subtends the same solid angle, although the number of star particles within them may vary. For a given radial shell, the probability of randomly selecting a star particle that falls into the $i^{th}$ pixel is given by,

\begin{equation}
p_i = \frac{n_i}{\sum_{i=1}^{N_{\text{pix}}} n_i},
\label{eq:probability}
\end{equation}

where $n_i$ is the number of star particles in the $i^{th}$ pixel. It is important to note that the mass of individual baryonic particles in TNG50 is fixed by the simulation’s resolution. However, each individual star particle is modeled as a single-burst simple stellar population (SSP) representing a collection of stars with the same birth time, metallicity, and mass. The total stellar mass associated with each star particle can differ due to variations in their formation history or the properties of the stellar populations they represent. To estimate star counts, we simply round the stellar mass of each star particle (in units of $M_\odot$) to the nearest whole number.

The information entropy associated with the random variable for a radial distance $r$ is given by,

\begin{equation}
H(r) = - \sum_{i=1}^{N_{\text{pix}}} p_i \log p_i.
\label{eq:information}
\end{equation}

The total number of star particles (mass-weighted) within the radial bin is given by,

\begin{equation}
\sum_{i=1}^{N_{\text{pix}}} n_i = N.
\label{eq:sum}
\end{equation}

Substituting \autoref{eq:probability} into \autoref{eq:information}, we obtain an equivalent expression for entropy:

\begin{equation}
H(r) = \log N - \frac{1}{N} \sum_{i=1}^{N_{\text{pix}}} n_i \log n_i.
\label{eq:info_entropy}
\end{equation}

For any given $r$, the maximum entropy $H_{\text{max}}$ occurs when the star distribution is perfectly isotropic, i.e., each pixel contains an equal number of star particles $n_i = \frac{N}{N_{\text{pix}}}$. In this case,

\begin{equation}
 H_{\text{max}} = \log N_{\text{pix}},
\label{eq:H_max}
\end{equation}

The whole-sky anisotropy parameter $a(r)$ is defined as the fractional deviation from the maximum entropy,

\begin{equation}
a(r) = 1 - \frac{H(r)}{H_{\text{max}}}.
\label{eq:anisotropy}
\end{equation}

This parameter quantifies the degree of anisotropy in the angular distribution of star particles within a given radial distance from the centre of the stellar halo. A value of $a(r) = 0$ corresponds to a perfectly isotropic (uniform) distribution, while higher values indicate increasing anisotropy.

\begin{figure*}
\resizebox{15cm}{13cm}{\rotatebox{0}{\includegraphics{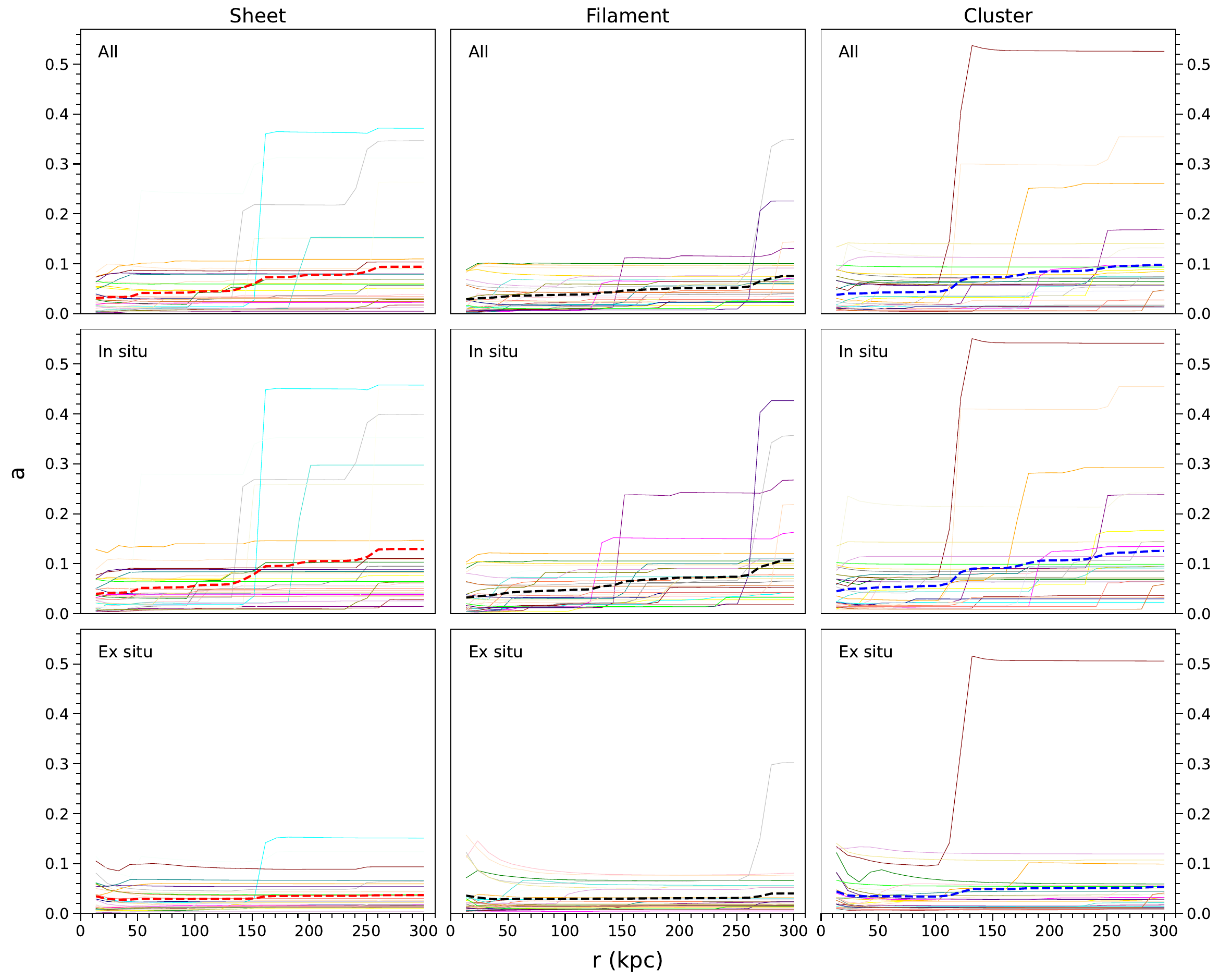}}}
\caption{The top three panels of this figure shows the radial variation of the whole-sky anisotropy in a set of stellar halos (29) residing in sheets, filaments and clusters respectively. The dotted line in each panel displays the mean whole-sky anisotropy of the 29 stellar halos in respective environment. The middle and bottom three panels show the results for the in situ and ex situ components of the same stellar halos. The same colour is used for the in situ, ex situ and combined components of each stellar halo within a given environment. We use $N_{\text{side}} = 32$ for the pixelization of the sky in each stellar halo in all environments.}
\label{fig:spatial_anis_combine}
\end{figure*}

\begin{figure*}
\resizebox{15cm}{5.0cm}{\rotatebox{0}{\includegraphics{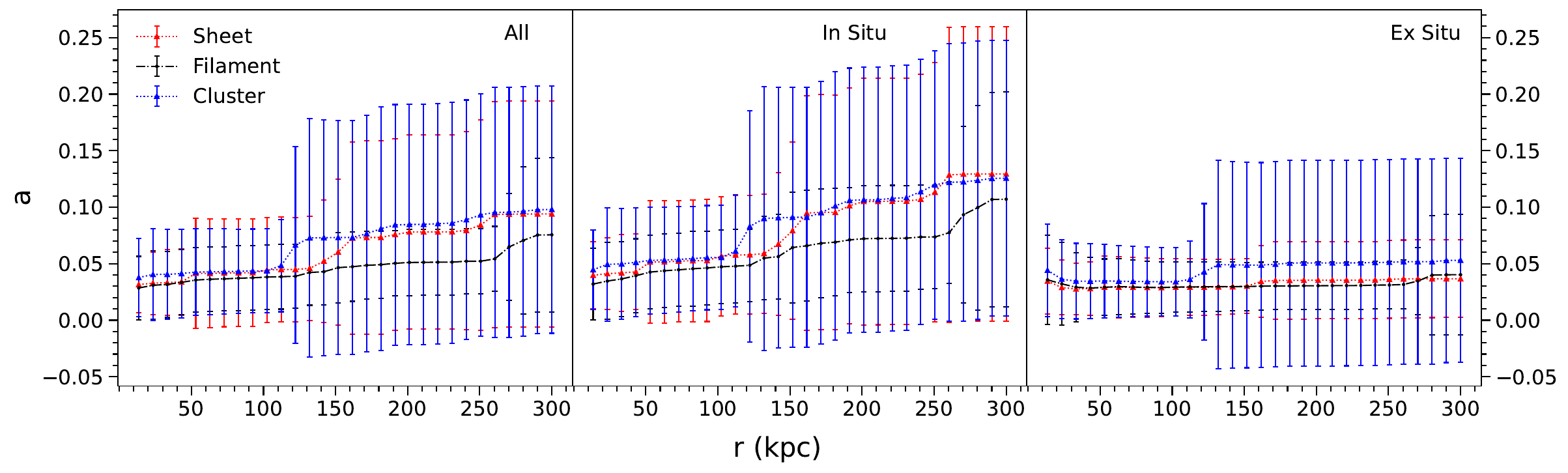}}}
\caption{The left panel of this figure compares the mean whole-sky anisotropy parameter as a function of radial distance from the centre of stellar halos residing in sheets, filaments, and clusters. The middle panel presents results for the in situ component, while the right panel shows those for the ex situ component of the stellar halo. The 1$\sigma$ error bars shown at each data point are computed using whole-sky anisotropy measurements of 29 halos from each environment.}
\label{fig:spatial_anis_meam}
\end{figure*}

\begin{figure*}
\resizebox{15cm}{13cm}{\rotatebox{0}{\includegraphics{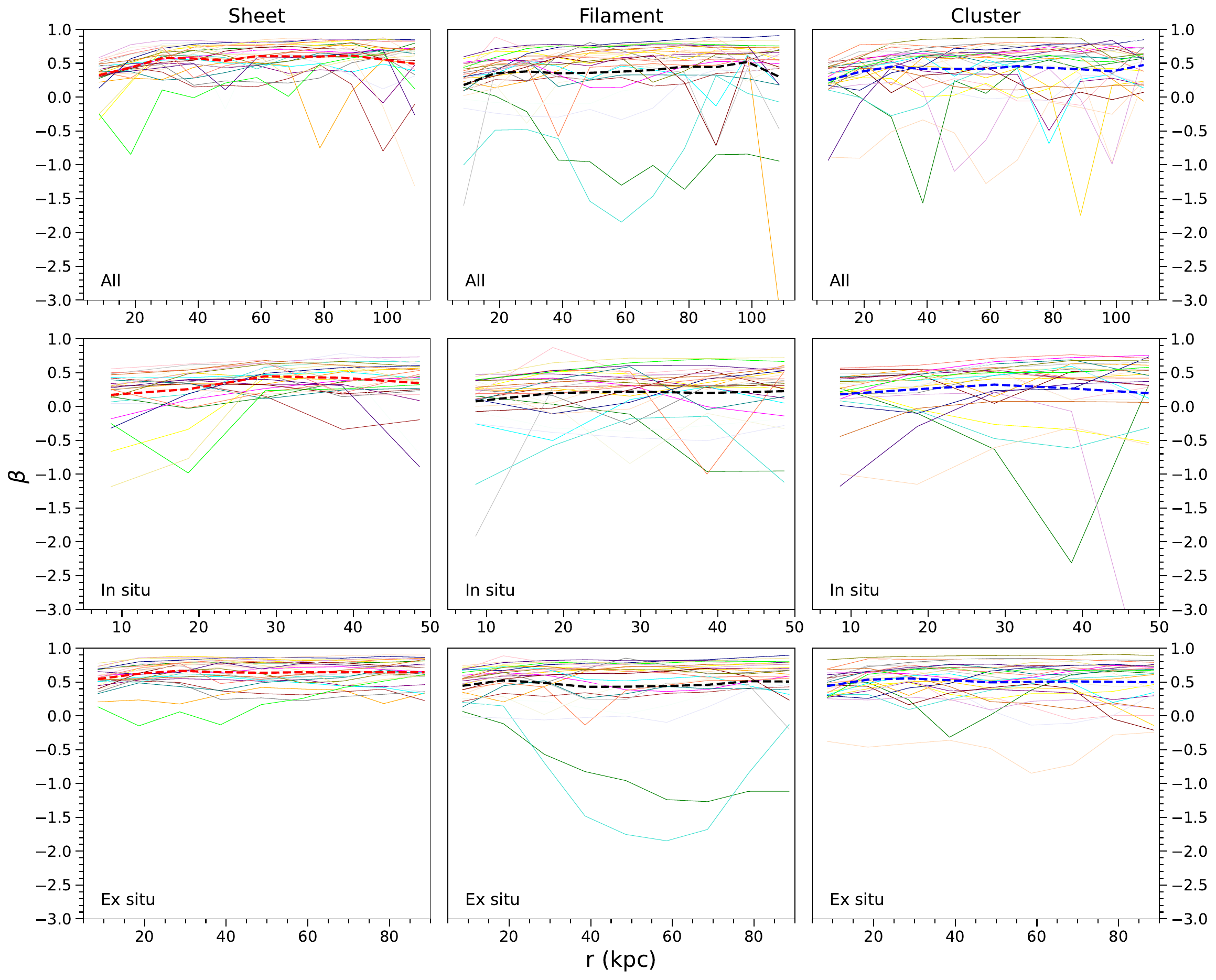}}}
\caption{The top three panels of this figure show the variation of the velocity anisotropy parameter ($\beta$) with radial distance from the centre of the 29 stellar halos in sheets, filaments and clusters respectively. The middle and bottom three panels show the same but for the in situ and ex situ components of stellar halo.}
\label{fig:velocity_anis_combine}
\end{figure*}

\begin{figure*}
\resizebox{15cm}{5.0cm}{\rotatebox{0}{\includegraphics{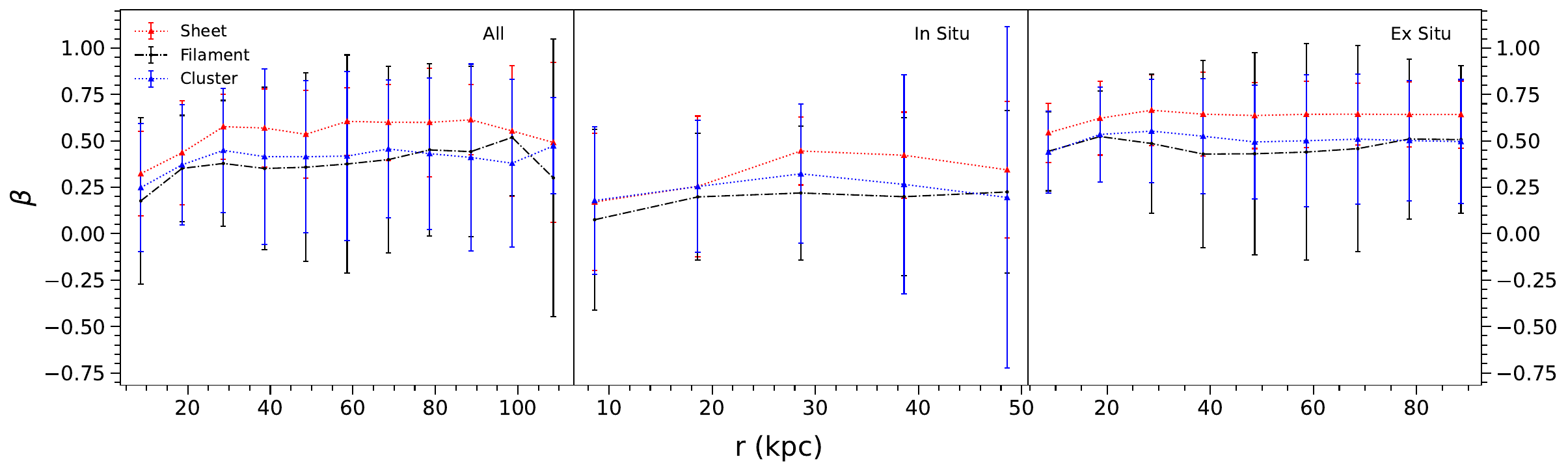}}}
\caption{The left panel compares the mean velocity anisotropy parameter $\beta$ as a function of radial distance from the centre of stellar halos residing in three different cosmic web environments (sheet, filament, cluster). The middle and right panels display the corresponding results for the in situ and ex situ components of the stellar halo, respectively. The 1$\sigma$ error bars shown at each data points are obtained using the velocity anisotropy measurements of 29 halos from each environment. }
\label{fig:velocity_anis_mean}
\end{figure*}

For a fixed stellar mass, the number of individual stars represented by a star particle varies with the age of the stellar population. Older star particles, which consist mostly of long-lived low-mass stars, represent more stars than younger particles, which include a larger fraction of short-lived massive stars. Although we do not account for the detailed distribution of stellar masses within each SSP, our method provides a simple and consistent proxy for relative star counts. We also perform the analysis for all star particles and for star particles younger than 10 Gyr in stellar halos across various web environments. The results are qualitatively similar in both cases, with some variations attributable to Poisson noise rather than systematic differences.

\subsubsection{Velocity anisotropy parameter }

To explore the kinematic structure of stellar halos within different cosmic web environments, we compute the velocity anisotropy parameter for 29 stellar halos from each cosmic web environment. The velocity anisotropy parameter $\beta$, originally introduced by \citet{binney80}, characterizes the orbital structure of stellar systems in spherical symmetry. It is a key component in Jeans dynamical modeling, where it is used to constrain the mass distribution of galaxies.

In a Galactocentric spherical coordinate system $(r, \theta, \phi)$, $\beta$ is defined as,

\begin{equation}	
\beta (r) = 1 - \frac{\sigma^2_\theta(r) + \sigma^2_\phi(r)}{2\sigma^2_r(r)},
\label{eq:velocity_anisotropy_parameter}
\end{equation}

where $\sigma_r$, $\sigma_\theta$, and $\sigma_\phi$ are the velocity dispersions in the radial and tangential ($\theta$, $\phi$) directions, respectively.

The parameter $\beta(r)$ ranges from $-\infty$ to 1. A value of $\beta = 1$ indicates purely radial orbits, while $\beta = -\infty$ corresponds to purely tangential motion. In general, positive values of $\beta$ imply radial orbital dominance, whereas negative values indicate a tangentially biased velocity distribution.

\section{Results}

We analyze 29 stellar halos from each of the three cosmic web environments (sheets, filaments, clusters) of the cosmic web and study the spatial anisotropy and the velocity anisotropy parameter as a function of halo-centric radius in each of them. 

The top three panels of \autoref{fig:spatial_anis_combine} illustrate the radial variation of the whole-sky anisotropy parameter for 29 stellar halos in each of the three cosmic web environments i.e. sheets, filaments, and clusters. In each panel, the dotted line represents the mean whole-sky anisotropy profile for the corresponding environment. We observe that, across all environments, whole-sky anisotropy tends to be lower in the inner regions of stellar halos and gradually increases with radial distance, reaching its peak in the outer halo. This trend suggests that the outer halo regions are more anisotropic, likely due to recent accretion of satellite material or infalling substructures. The step-like jumps in the anisotropy profiles, prominently visible in several halos are indicative of bound substructures contributing anisotropically at specific radial shells. These results are consistent with earlier findings by \citet{pandey22} and \citet{mondal24}. Despite being in the same type of cosmic web environment, individual stellar halos exhibit a wide range of whole-sky anisotropy values, highlighting the inherently stochastic and diverse nature of their assembly histories \citep{bullock05, cooper10, monachesi19, harmsen17, merritt16, merritt20, gozman23}. To further investigate the role of cosmic web environment on the formation and structure of stellar halos, we separate the stars into two components: in situ and ex situ. We then analyze the spatial and velocity anisotropies of these components of the 29 Milky Way-mass galaxies drawn from each of the three cosmic web environments. We show the results for the in situ and ex situ stars in the middle and bottom three panels of \autoref{fig:spatial_anis_combine} respectively. Interestingly, despite expectations that in situ stars would form smoother and more isotropic structures, we observe more step-like features in their whole-sky anisotropy profiles compared to the ex situ component. This may reflect the cumulative impact of episodic disk heating, merger-induced perturbations, or bursty star formation that redistribute in situ stars into coherent shells \citep{villa08, purcell10, jean17}, while ex situ stars accreted through more continuous, diffuse infall tend to exhibit smoother radial distributions.

We compare the mean whole-sky anisotropy profiles in sheets, filaments, and clusters in the three panels of \autoref{fig:spatial_anis_meam}. The results for all, in situ and ex situ stars indicate that, on average, stellar halos in sheets and clusters exhibit a somewhat higher whole-sky anisotropies than those in filaments. However, these differences are not statistically significant due to the large 1$\sigma$ errorbars (particularly in sheets and clusters) making it difficult to draw firm conclusions. The 1$\sigma$ errorbars at a given radius represent the standard deviation across the 29 stellar halos in each environment, thus quantifying the halo-to-halo variation.

It is important to note that the computed whole-sky anisotropy is sensitive to the angular resolution used to pixelate the sky. As shown in \citet{pandey22} and \citet{mondal24}, the choice of the HEALPix resolution parameter $N_{\text{side}}$ can influence the measured anisotropy values by affecting the granularity of angular bins. In this work, we adopt a consistent resolution of $N_{\text{side}} = 32$ across all halos to ensure uniformity in our analysis and minimize resolution-induced biases.

The top three panels of \autoref{fig:velocity_anis_combine} show the radial profiles of the velocity anisotropy parameter $\beta(r)$ for stellar halos in sheets, filaments, and clusters. In each panel, the dotted line denotes the mean velocity anisotropy profile computed from 29 stellar halos within the respective cosmic web environment. The individual $\beta(r)$ profiles show considerable complexity, with multiple peaks and troughs evident in all environments. These oscillations reflect the intricate kinematic structure of stellar halos and are often associated with the presence of bound substructures at specific radii. Notably, dips in the $\beta$ profiles frequently correspond to infall events or satellite debris that retain coherent tangential motion, while negative $\beta$ values at certain radii point to regions where tangential or rotationally supported orbits dominate. Despite these local variations, the overall trend in each environment indicates a dominance of radial orbits across most of the radial extent of the stellar halos, consistent with expectations from hierarchical structure formation where radial accretion events are common. However, the presence of sharp deviations from this trend underscores the impact of recent or ongoing mergers, anisotropic mass accretion, and the dynamical mixing processes that influence halo kinematics. As with whole-sky anisotropy, the velocity anisotropy parameter also exhibits significant scatter among halos within the same cosmic web environment. This reinforces the notion that the assembly history of stellar halos is highly stochastic, shaped not only by the environment but also by the diversity of local accretion events and orbital dynamics even for halos of similar mass. We also show the velocity anisotropy of the in situ and ex situ stars as a function of halo-centric distance in the middle and bottom three panels of \autoref{fig:velocity_anis_combine}. We compute and compare the mean $\beta(r)$ profiles for all, in situ and ex situ stars in stellar halos in sheets, filaments, and clusters, in the three panels of \autoref{fig:velocity_anis_mean}. The average profiles suggest that halos in sheets and clusters tend to exhibit slightly higher velocity anisotropy compared to those in filaments, nearly at all radii. However, due to the large $1\sigma$ uncertainties, these apparent differences are not statistically robust and should be interpreted with caution.

It is important to note that the radial extent of the whole-sky anisotropy and velocity anisotropy profiles in stellar halos is not the same. The whole-sky anisotropy, computed using integrated stellar number counts across the full sky, can be measured out to the outermost boundaries of the halo. In contrast, the velocity anisotropy parameter $\beta$ is calculated within discrete, non-overlapping radial bins and is more limited in range. This is because the stellar particle number density decreases with radius, reducing the statistical reliability of $\beta$ at large distances. In our analysis, we compute $\beta$ only in radial bins containing at least 100 stellar particles to ensure robust statistics. As a result, the $\beta$ profiles typically extend to smaller radii compared to the whole-sky anisotropy profiles.

\section{Conclusions}

Understanding the influence of large-scale cosmic environment on the outer stellar components of galaxies namely, their stellar halos is crucial for building a comprehensive framework of galaxy formation and evolution. In this study, we investigated whether the geometric environment of the cosmic web leaves a detectable imprint on the spatial and kinematic structure of stellar halos in Milky Way-mass galaxies, using high-resolution data from the TNG50 simulation. Contrary to intuitive expectations and previous studies linking galaxy properties to environment, our analysis finds no statistically significant difference in either spatial or velocity anisotropy of stellar halos across different cosmic web environments. This suggests that, at fixed halo mass, the influence of the large-scale environment on the structure and dynamics of stellar halos is either negligible or too subtle to detect with the current dataset. Our results imply that the assembly of stellar halos is likely governed by highly stochastic and small-scale processes such as the number, mass, orbital configuration, and timing of satellite mergers rather than the global geometry of the cosmic web. While the cosmic web plays a key role in modulating the inflow of matter, its impact on the fine-grained kinematic and spatial features of stellar halos appears weak or non-deterministic in this mass regime.

Our findings suggest that the influence of the cosmic web on the internal kinematic and structural properties of stellar halos is minimal, regardless of the origin of the stellar material. The lack of environmental dependence in the ex situ stars is particularly notable, as these stars are assembled through satellite accretion-processes often thought to be modulated by large-scale anisotropic infall along filaments. It is important to note, however, that a null result does not necessarily imply the complete absence of environmental influence. The interplay between large-scale environment and the localized accretion and merger processes may be more complex. The inherently chaotic nature of halo assembly may wash out weak correlations with environment, especially when averaged over many stochastic merger histories. Furthermore, the dominant role of halo mass in governing stellar halo properties may overshadow any subtle environmental trends. At fixed mass, galaxies in different environments might undergo similar types of accretion events, leading to convergence in their outer structures. Additional factors may also contribute to the lack of detectable signal. Feedback processes, baryonic physics, and the survival and disruption of infalling satellites can all affect the morphology and kinematics of stellar halos in complex, non-linear ways. Moreover, the finite resolution of the TNG50-1 simulation may limit our ability to capture the full diversity of satellite merger events, particularly in the outer halo where kinematic features are most sensitive to environmental anisotropy. Our analysis reinforces the picture of stellar halo assembly as a complex, highly non-deterministic process, only weakly constrained by environment at fixed mass. It highlights the importance of stochasticity in stellar halo assembly and the need for caution when attributing halo properties to cosmic environment \citep{tojeiro17, ramakrishnan19, hunde25}.

A caveat of our analysis is that each star particle in the simulation represents an entire single stellar population (SSP) with an assumed Initial Mass Function. For SSPs with the same initial stellar mass, the number of individual stars they correspond to varies with age. Older SSPs contain a larger population of low-mass, long-lived stars, whereas younger SSPs still retain a higher fraction of massive, short-lived stars. Consequently, the mass of a star particle cannot be directly converted into an exact number of individual stars. It may be noted that we use the stellar mass (in units of $M_{\odot}$), rounded to the nearest integer, as a proxy for the number of stars in an SSP. To test the robustness of our results under this approximation, we repeat the anisotropy analyses using only SSPs younger than 5 Gyr. This additional test confirms that our main conclusions remain unchanged.

The environmental imprint on the spatial and kinematic structure of stellar halo may become more pronounced at higher masses. More massive halos may experience anisotropic accretion more coherently, particularly along filaments \citep{veena18, veena19, kraljic20}, potentially imprinting stronger spatial and kinematic signatures on their stellar halos. Previous works \citep{pillepich18a, pillepich18b} have found evidence for more extended and structured stellar halos in high-mass galaxies located in denser environments. However, such galaxies are relatively rare within the modest volume of the TNG50 simulation ($51.7^3 \, \mathrm{Mpc}^3$), and any statistical analysis of them is likely limited by small number statistics and cosmic variance. To robustly assess environmental dependencies across a broad range of galaxy masses and environments, future studies will require high-resolution simulations with significantly larger cosmological volumes. These will enable the detection of subtle trends, reduce statistical uncertainties, and allow for detailed modeling of both the stochastic and deterministic components of stellar halo formation.

\section{Acknowledgement}
We thank the scientific editor Prof. Joop Schaye and an anonymous reviewer for insightful comments and constructive suggestions that greatly improved this paper. AM thanks Andrew Cooper, Analisa Pillepich and Diego Sotillo Ramos for useful discussions. AM acknowledges UGC, Government of India for support through a Junior Research Fellowship. BP acknowledges financial support from the SERB, DST, Government of India through the project CRG/2019/001110. BP also acknowledges IUCAA, Pune, for providing support through the associateship programme. AN thanks Dylan Nelson for help in understanding the IllustrisTNG data. AN also acknowledges the financial support from the Department of Science and Technology (DST), Government of India through an INSPIRE fellowship. 

\section{Data availability}
The IllustrisTNG simulation data used in this work are publicly available at \url{https://www.tng-project.org/}. The data produced in this study are available from the authors upon reasonable request.

% Don't change these lines
\bsp	% typesetting comment
\label{lastpage}
\end{document}